\documentclass[11pt]{article}
\usepackage{amsmath}
\usepackage{pstricks}
\usepackage[vcentermath]{youngtab}
\usepackage{bbm}
\usepackage{amssymb,amsfonts}

\newcommand{\re}[1] {(\ref{#1})}

\def\half{\frac{1}{2}}
\def\bsh{\backslash}

\newfont{\bbbold}{msbm10 scaled \magstep1}

\def\bbC{\mbox{\bbbold C}}

\def\cA{{\cal A}}
\def\cB{{\cal B}}
\def\cC{{\cal C}}
\def\cD{{\cal D}}

\def\cL{{\cal L}}
\def\cM{{\cal M}}
\def\cN{{\cal N}}
\def\cO{{\cal O}}

\def\cU{{\cal U}}

\newfont{\goth}{eufm10 scaled \magstep1}

\def\gg{\mbox{\goth g}}

\def\gl{\mbox{\goth l}}

\def\go{\mbox{\goth o}}
\def\gp{\mbox{\goth p}}

\def\gs{\mbox{\goth s}}

\def\gu{\mbox{\goth u}}

\def\a{\alpha}\def\adt{\dot \alpha}
\def\b{\beta}\def\bdt{\dot \beta}
\def\c{\gamma}\def\C{\Gamma}
\def\d{\delta}\def\D{\Delta}
\def\e{\epsilon}\def\ve{\varepsilon}
\def\f{\phi}

\def\k{\kappa}
\def\l{\lambda}

\def\th{\theta}

\def\be{\begin{equation}}\def\ee{\end{equation}}
\def\bea{\begin{eqnarray}}\def\eea{\end{eqnarray}}
\def\barr{\begin{array}}\def\earr{\end{array}}
\def\x{\xi}
\def\o{\omega}
\def\del{\partial}

\def\xz{\times}

\def\nab{\nabla}

\let\la=\label

\def\nn{\nonumber}
\def\bd{\begin{document}}
\def\ed{\end{document}}
\def\ba{\begin{array}}
\def\ea{\end{array}}
\def\bea{\begin{eqnarray}}
\def\eea{\end{eqnarray}}
\def\ft#1#2{\tfrac{#1}{#2}}
\def\fft#1#2{\frac{#1}{#2}}
\def\sst#1{{\scriptscriptstyle #1}}
\def\oneone{\rlap 1\mkern4mu{\rm l}}

\newcommand{\eq}[1]{(\ref{#1})}
\newcommand{\w}[1]{\\[0.#1cm]}
\def\eqs#1#2{(\ref{#1}-\ref{#2})}
\def\det{{\rm det\,}}
\def\tr{{\rm tr}}
\def\ad{{\rm ad}}

\newcommand{\hoch}[1]{$\, ^{#1}$}
\newcommand{\imperial}{\it\small Theoretical Physics Group, Imperial College London\\ Prince Consort Road, London SW7 2AZ, UK}
\newcommand{\kings}
{\it\small Department of Mathematics, King's College, University of London\\ Strand, London WC2R 2LS, UK}
\newcommand{\uu}
{\it\small Department of Theoretical Physics, Uppsala, Sweden}
\newcommand{\hip}
{\it\small HIP-Helsinki Institute of Physics, P.O. Box 64 FIN-00014
University of Helsinki, Suomi-Finland}
\newcommand{\stock}
{\it\small Department of Theoretical Physics, Stockholm, Sweden}
\newcommand{\golm}
{\it\small AEI, Max Planck Institut f\"ur Gravitationsphysik\\ Am M\"{u}hlenberg 1, D-14476 Potsdam, Germany}
\makeatletter
\renewcommand\theequation{\thesection.\arabic{equation}}
\@addtoreset{equation}{section} \makeatother

\newcommand{\sa}{/ \hspace{-1.2ex}}
\newcommand{\saa}{/ \hspace{-1.4ex}}
\newcommand{\saaa}{\, / \hspace{-1.6ex}}
\newcommand{\Scal}[1]{\Bigl ({#1} \Bigr )}
\newcommand{\scal}[1]{\bigl ({#1} \bigr )}

\newcommand{\CR}{\nonumber \\*}

\newcommand{\trace}{\hbox {tr}~}
\newcommand{\traceS}{\hbox {tr}_{\scriptscriptstyle \mathfrak{S}}~}

\DeclareMathAlphabet{\mathpzc}{OT1}{pzc}{m}{it}
\def\BRST{\,\mathpzc{s}\,}
\def\aBRST{{\scriptstyle (\mathpzc{s})}}
\def\q{{{\scriptscriptstyle (Q)}}}
\def\qs{{\scriptscriptstyle (Q\mathpzc{s})}}
\def\Qsla{{\mathcal{S}_{\q}}}
\def\Slav{{\mathcal{S}_\aBRST}}
\def\epsilonb{{\overline{\epsilon}}}
\def\bulletup{{\scriptstyle \bullet}}

\newcommand{\gra}[2]{{\scriptscriptstyle (#1 , #2 )}}
\newcommand{\ord}[1]{{\scriptscriptstyle (#1)}}

\def\cL{{\cal L}}
\def\cN{\mathcal{N}}
\def\cO{\mathcal{O}}

\def\ie{{\it i.e.}\ }
\def\eg{{\it e.g.}\ }

\newcommand{\sfrac}[2]{{\scriptstyle \frac{#1}{#2}}}
\newcommand{\stfrac}[2]{{\scriptscriptstyle \frac{#1}{#2}}}

 \def\balpha{{\overline{\alpha}}}
 \def\bbeta{{\overline{\beta}}}
 \def\bgamma{{\overline{\gamma}}}
 \def\bdelta{{\overline{\delta}}}
 \def\bepsilon{{\overline{\epsilon}}}
 \def\bvarepsilon{{\overline{\varepsilon}}}
 \def\bzeta{{\overline{\zeta}}}
 \def\bareta{{\overline{\eta}}}
 \def\btheta{{\overline{\theta}}}
 \def\bvartheta{{\overline{\vartheta}}}
 \def\biota{{\overline{\iota}}}
 \def\bkappa{{\overline{\kappa}}}
 \def\blambda{{\overline{\lambda}}}
 \def\bmu{{\overline{\mu}}}
 \def\bnu{{\overline{\nu}}}
 \def\bxi{{\overline{\xi}}}
 \def\bpi{{\overline{\pi}}}
 \def\brho{{\overline{\rho}}}
 \def\bvarrho{{\overline{\varrho}}}
 \def\bsigma{{\overline{\sigma}}}
 \def\bvarsigma{{\overline{\varsigma}}}
 \def\btau{{\overline{\tau}}}
 \def\bphi{{\overline{\phi}}}
 \def\bvarphi{{\overline{\varphi}}}
 \def\bchi{{\overline{\chi}}}
 \def\bpsi{{\overline{\psi}}}
 \def\bomega{{\overline{\omega}}}

\def\thalf{{\textrm{\tiny\textonehalf}}}
\def\tquarter{{\textrm{\tiny\textonequarter}}}
\def\Ko{{\scriptscriptstyle K}}
\def\tKo{\scriptscriptstyle k }
\def\corr{$\clubsuit$}

\newcommand{\auth}{\large P.S.\ Howe${}^{a,}$\footnote{email: paul.howe@kcl.ac.uk} and U. Lindstr\"om${}^{b,c,}$\footnote{email: ulf.lindstrom@physics.uu.se}}

\begin{document}

\renewcommand{\thefootnote}{\fnsymbol{footnote}}

\null
\begin{flushright}
{\small KCL-MTH-16-10}\\
{\small UUITP-35/16}\\
{\small Imperial-TP-UL-2016-02}\\
\vskip 1.5 cm
\end{flushright}

\begin{center}
{\Large{\bf Super-Laplacians and their symmetries}}
\vspace{.75cm}

\auth
\end{center}
\vspace{.5cm}

\centerline{${}^a${\it \small Department of Mathematics, King's College London}}
\centerline{{\it \small The Strand, London WC2R 2LS, UK}}
\vspace{.5cm}
\centerline{${}^b${\it \small Department of Physics and Astronomy, Theoretical Physics, Uppsala University}}
\centerline{{\it \small SE-751 20 Uppsala, Sweden }}
\vspace{.5cm}
\centerline{${}^c${\it \small Theoretical Physics, Imperial College, London}}
\centerline{{\it \small Prince Consort Road, London SW7 2AZ, UK}}

\vspace{1cm}


\centerline{{\bf Abstract}}
\vskip .5cm
A super-Laplacian is a set of differential operators in superspace whose highest-dimensional component is given by the spacetime Laplacian. Symmetries of super-Laplacians are given by linear differential operators of arbitrary finite degree and are determined by superconformal Killing tensors. We investigate these in flat superspaces. The differential operators determining the symmetries give rise to algebras which can be identified in many cases with the tensor algebras of the relevant superconformal Lie algebras modulo certain ideals. They have  applications to Higher Spin theories.

\vspace{1cm}


\renewcommand{\thefootnote}{\arabic{footnote}}
\setcounter{footnote}{0}

\pagebreak
\tableofcontents
\setcounter{page}{1}


\section{Introduction}

In \cite{Eastwood:2002su} a generalised symmetry of the Laplacian was defined to be a linear differential operator $\cD$ that preserves the Laplacian $\D$ in the sense that
\be
\D\cD=\d \D
\la{1.1}
\ee
where $\d$ is another linear differential operator. Such a differential operator therefore maps solutions of the Laplace equation to new ones. It was further shown that such symmetries are determined by conformal Killing tensors (CKTs), and that for a given CKT, corresponding to the highest-degree term in derivatives,  there are canonical choices of the lower-degree terms and of $\d$. The analysis of such symmetries was carried out in full detail in Euclidean spaces of arbitrary dimension, and it was shown that these symmetries, defined modulo the Laplacian itself, determine an associative algebra. Furthermore, this algebra can be described in terms of the tensor algebra of the conformal Lie algebra, $\gg$, modulo a certain ideal, the Joseph ideal, which had been introduced previously in \cite{Joseph:76}. Alternatively, it can be described in terms of the universal enveloping algebra $\cU_{\gg}$. This algebra has a physical application in higher-spin theory  via the AdS/CFT correspondence (for a relevant review, see \cite{Boulanger:2013zza}). 

In this paper we shall attempt to generalise Eastwood's construction to the supersymmetric case.\footnote{ In a different context, Killing tensor superfields were discussed in relation to an infinite-dimensional superalgebra in $D=4, N=1$ anti-de Sitter superspace in \cite{Gates:1996xs}, based on the earlier \cite{Gates:1996my}.} It is not possible to follow his construction directly for the following reasons: firstly, one cannot present super Minkowski space in terms of a higher-dimensional ambient super Minkowski space   with two extra even dimensions (one time-like)\footnote{It is possible to make embeddings into other superspaces, however, see \cite{Kuzenko:2012tb}.}; secondly, the Laplacian itself has to be changed, and thirdly, one has to be careful in extending Lie-algebraic 
results to the super case due to the occurrence of reducible but indecomposable representations which arise even in the  finite-dimensional   case, see \cite{Bars:1982se,Leites:2002} and references therein. It is also the case that there are superconformal algebras in flat superspaces only for $D=3,4\,\&6$, for arbitrary numbers, $N$, of supersymmetries and in $D=5$  for $N=1$ supersymmetry \cite{Nahm:1977tg}. In the even case, the Laplacian (or d'Alembertian in spacetime) is taken to act on scalar fields which must have the right, dimension-dependent, conformal weight if we require conformal covariance. In the super case one needs to consider superconformal massless supermultiplets, which are special cases of irreducible unitary representations of superconformal groups \cite{Dobrev:1985qv}. We shall concentrate in this paper on the minimal multiplets which are the analogue of conformal scalar fields.\footnote{One might expect symmetries of more general multiplets to involve other tensors in addition to those of Killing type; in the non-supersymmetric case, for example, it is known that the Dirac operator has symmetries related to Killing-Yano tensors, see \eg \cite{Gibbons:1993ap}\black} We remark in passing that there is a more straightforward extension of  \cite{Eastwood:2002su} in super Euclidean space \cite{css}, but the discussion given there is not directly relevant to the case of supersymmetric field theory, although it does have applications to the algebraic structure given by the  higher-spin \black symmetries.

The first task is to define an appropriate notion of a super-Laplacian. Some partial results for $N=1$ supersymmetry in $D=3,4$ have been reported in \cite{jpm}, but to our knowledge, this has not been extended to the general case. The problem is that the Laplacian itself does not suffice, one also needs to consider fermionic differential operators such as the Dirac operator, and further operators in the case of extended supersymmetry. In Minkowski superspace one must identify the multiplets in terms of constrained superfields and then determine the relevant super-Laplacians in each case.\footnote{ Laplacian might seem an inappropriate term for multiplets without scalars, whose components typically have first-order component field equations, but it is still the case that these satisfy the wave equation.\black} A simple example is given by a scalar multiplet in $N=1, D=3$ supersymmetry, where one could regard the operator $D^2=D_\a D^\a$ to be the Laplacian  ($D_\a=\frac{\del}{\del\th^\a}+\frac{i}{2}(\c^a)_{\a\b}\del_a$, where the (3 even, 2 odd) coordinates are $(x^a,\th^\a)$). \black However, even in this case one would need to take into account the higher  components of $D^2$, \ie $ (\c^a)^{\a\b} \del_a D_\b$ and $\D$, and it is in this sense that we mean a set of operators including the spacetime Laplacian. For $D=4$ super Minkowski space,  the unitary massless superconformal superfields for arbitrary spins were given for $D=4$ some time ago \cite{Siegel:1980bp,Howe:1981qj},  and for all dimensions in \cite{Siegel:1999ew}\black. But there are other possible superspaces which can also be defined locally as cosets of the superconformal groups, such as harmonic superspaces \cite{Rosly,Galperin:1984av,Karlhede:1984vr}. We shall focus on the minimal multiplets formulated in a particular class of harmonic superspaces,  known as analytic superspaces, because in this case these multiplets admit local descriptions as single-component fields that are annihilated by sets of differential operators that can be written as tensorial products of two derivatives. The minimal multiplets are those whose maximal spin (or helicity) components are the lowest possible for each value of $D$ and $N$. They have been considered from a somewhat different point of view in the supersymmetric higher-spin context in \cite{Govil:2013uta,Govil:2014uwa,Fernando:2015tiu}.

It  turns out that life is simpler in some ways in analytic superspace, at least if one restricts oneself to the minimal multiplets. The super-Laplacians acting on these multiplets are second-order differential operators whose highest-dimensional components are the spacetime Laplacians. Analytic superspaces have fewer odd co-ordinates than the corresponding Minkowski superspaces but they also have internal coordinates associated with cosets of the internal (R)-symmetry groups.  We discuss these operators in section 2 after a brief review of analytic superspace and superconformal  Killing tensors (SCKTs). We also take the opportunity to discuss currents in this setting. In section 3 we study the linear differential operators that give rise to symmetries of super Laplacians in analytic superspace. Since a super-Laplacian is a set of operators $\D_I$, the notion of a symmetry must be amended to
\be
\D_I \cD=d_I{}^J \D_J\ ,
\la{1.2}
\ee
where $d_I{}^J$ denotes another set of differential operators. In a similar fashion to the bosonic case any such $\cD$ is determined by a SCKT; we show that an $n$th-order differential operator $\cD_K$ has a leading term determined by an $n$th-rank SCKT $K$ and that all of the lower-order terms in $\cD_K$ can be constructed in a systematic fashion. The analytic superspace formalism allows one to treat all values of $N$  for each $D$ at the same time. For $D=3,4$, for ease of presentation, we shall focus on the case of $N=2M$, although we also briefly remark on SCKTs in $D=4$ for $N=2M+1$.

It is clear from \eq{1.2} that such symmetries will give rise to an associative algebra. In section 4 we analyse products of two operators determined by superconformal Killing vectors, $K,L$ and show that, in most cases,
\be
\cD_K \cD_L=\cD_{K\circledcirc L} + \half \cD_{[K,L]} +  K\cdot L\ ,
\la{1.3}
\ee
where $K\circledcirc L$ denotes the highest-weight representation in the product of $K,L$, $[K,L]$ the commutator in the appropriate Lie superalgebra $\gg$,  and the last term is a scalar function. Since the last term must also be a symmetry of the super-Laplacian it follows that it must be a constant. 

The above equation determines the Joseph ideal \cite{Joseph:76} when it is well-defined, and thus leads on in a  natural way to the discussion of the symmetry algebras associated with these differential operators that we give in section 5. But we can also approach this more directly in super-twistor space for the case $D=4$, following the treatment of the series of complex Lie algebras $\gs\gl(n)$ given in \cite{Eastwood:2007}. This is done in section 6. There is a short appendix on our sign conventions.

\section{Super-Laplacians in analytic superspaces}
\subsection{Analytic superspace}
We collect here the basics about analytic superspaces in $D=3,4\&6$.\footnote{Analytic superspaces are harmonic superspaces which have fewer odd coordinates than conventional Minkowski superspaces. We shall use the local super-coordinate formulation of \cite{Howe:1995md}.} These superspaces are (open subsets of) cosets of the relevant complexified superconformal groups, $SpO(2|N), SL(4|N)$ and $OSp(8|N)$ for $D=3,4\&6$ respectively, that resemble (and extend) the corresponding complex spacetimes in the spinor formalism, for which coordinates carry a pair of spinor indices. The superconformal groups act linearly on super-twistor spaces which are $\bbC^{4|N}$ for $D=3,4$ and $\bbC^{8|2N}$ for $D=6$. The spaces we are interested in are spaces of planes that have half the dimensions of super-twistor spaces, so planes of super-dimension $(2|M)$ for $D=3,4$, where $N=2M$, and $(4|N)$ for $D=6$. These spaces are (maximal) super-Grassmmannians and in $D=3,6$ there are also required to be isotropic with respect to the super-symplectic form for $D=3$ or the ortho-symplectic metric for $D=6$. 

In all cases we can define local maps from analytic superspace $\cM_A$ to the relevant supergroup by
\be
\cM_A\ni X \rightarrow \left( \barr{cc} 1 & X\\ 0 & 1 \earr\right)\ .
\la{2.1}
\ee
The coordinate matrices $X$ have indices as follows:
\begin{align}
D=3: X^{AB}&=X^{BA}=(x^{\a\b},\x^{\a b}, y^{ab})\nn\w1
D=4: X^{AA'}&=(x^{\a\a'},\x^{\a a'}, \x^{a\a'}, y^{aa'})\nn\w1
D=6: X^{AB}&=-X^{BA}=(x^{\a\b},\x^{\a b}, y^{ab})\ ,
\label{2.2}
\end{align}
where $x$ are the spacetime coordinates, $y$ are the internal even coordinates and $\x$ the odd coordinates. The (anti)-symmetry for $D=6,3$ is understood to be graded and we regard $\a,\a'$ as even indices and $a,a'$ as odd ones. The $y$s parameterise internal coset spaces of the R-symmetry groups, for example, in $D=4$, these will be of the form $(U(M)\xz U(M))\backslash U(2M)$. There are therefore additional even coordinates as compared to conventional superspace but there are only half the number of odd ones. Since these spaces are cosets it is easy to compute the action of the conformal Lie superalgebras on the coordinates; we find
\be
\d X=b + a X + X d + X c X\ ,
\la{2.3}
\ee
corresponding to the element
\be
z=\left( \barr{cc} -a & b\\ -c& d \earr\right)
\la{2.4}
\ee
in the superconformal algebra $\gg$. In the $D=3,6$ cases the matrix parameters $a\  \&\ d$ are related to each other by transposition. Note that the above formalism remains applicable when $N=0$, in which case $X$ reduces to $x$, and $\gg$ to the corresponding conformal algebra.

\subsection{Currents and superconformal Killing tensors}
In Minkowski superspace supercurrents are defined in different ways depending on the theory under consideration and the number of supersymmetries. However, in analytic superspaces we can define currents in a similar way to ordinary spacetime. A current in $D=3,4$ or $6$ spacetime in spinor notation is $J_{\a\b}$, (anti)-symmetric for $D=6,3$, (or $J_{\a\a'}$ $D=4$), and the conservation condition is simply
\be
\del^{\a\b} J_{\a\b}=0,\ D=3,6\qquad {\rm or} \qquad\del^{\a\a'} J_{\a\a'}=0, \ D=4\ .
\la{2.5}
\ee
In analytic superspace we do not have an invariant metric that can be used to raise or lower indices so these formulae cannot be taken over unchanged. However, it is easy to rewrite them without a metric:
\begin{align}
D&=3: \ \del_{[\a[\c} J_{\b]\d]}=0 \ ,\nn\w1
D&=4:\del_{[\a[\a'} J_{\b]\b']}=0\ ,\nn\w1
D&=6: \ \ \ \del_{[\a\b} J_{\c\d]}=0\ ,
\label{2.6}
\end{align}
where in the $D=3$ the anti-symmetrisation is over the $\a\b$ and $\c\d$ pairs separately. These formulae can be taken over straightforwardly to analytic superspace. A conserved current can be defined to be a co-vector $J_{AB}$ or $J_{AA'}$ satisfying
\begin{align}
D&=3: \ \del_{[A[C} J_{B]D]}=0 \ ,\nn\w1
D&=4:\del_{[A[A'} J_{B]B']}=0\ ,\nn\w1
D&=6: \ \ \, \del_{[AB} J_{CD]}=0 \ .
\label{2.7}
\end{align}
We can also extend these definitions to graded-symmetric currents of arbitrary degree $n$ (which have $2n$ indices in this formalism). These are given by tensors $J_{A_1\ldots A_{2n}}$, totally graded symmetric, for $D=3$, tensors of the form $J_{A_1\ldots A_n,A'_1\ldots A'_n}$, totally graded-symmetric on both  the unprimed and primed sets of indices, for $D=4$, and, for $D=6$, tensors of the form $J_{A_1 A_2,A_3 A_4,\ldots ,A_{2n-1} A_{2n}}$, graded-antisymmetric on each pair, graded-symmetric under the interchange of any two pairs, and such that graded anti-symmetrisation on any three indices gives zero. For each of these, a conservation condition as in \eq{2.7} is defined for any index pair. For example, for $D=3$ a rank $n$ current is given by a totally graded-symmetric tensor satisfying
\be
\del_{[A[B} J_{C]D]E_3\ldots E_{2n}}=0\ .
\la{2.8}
\ee

As an example we consider the set of higher-spin currents in $N=4, D=4$ super Yang-Mills theory which were constructed in \cite{Bianchi:2005ze}. The basic field is the minimal multiplet which we discuss below. It is a single-component analytic superfield $W$ satisfying the  super-Laplace equation $\del_{[A[A'}\del_{B]B']}W=0$ and the currents are obtained by sandwiching derivatives between two such fields, in a similar manner to the construction given for free bosonic scalars in \cite{Mikhailov:2002bp}. The currents are
\be
J_{A_1\ldots A_{n+2},A'_1\ldots A'_{n+2}}=
\sum_{k=0}^{n+2}(-1)^k \binom{n+2}{k}^2 \del^k_{(A_1\ldots A_k(A'_1\ldots A'_k}W\del^{n+2-k}_{A_{k+1}\ldots A_{n+2})A'_{k+1}\ldots A'_{n+2})}W\ .
\la{2.8.1}
\ee
They are conserved due to the fact that the super-Laplacian annihilates $W$. (In \cite{Bianchi:2005ze} this formula was given for the (non-interacting) non-Abelian case where one has to take the trace over the Yang-Mills indices on $W$.) Note that for $n$ odd this expression gives total derivatives. The lowest possible current is actually when $n=-2$, \ie no derivatives; this is the energy-momentum supermultiplet, while for $n=0$ one finds the Konishi multiplet.

The currents can be thought of  as being dual in some sense to the superconformal Killing tensors (SCKTs). These are contravariant tensors with similar index structures obeying the following constraints \cite{Howe:2015bdd}: 
\subsubsection*{$D=3$}
\be
\del_{A_1 A_2} K^{B_1\ldots B_{2n}}= a_n\, \d_{(A_1}{}^{(B_1} \del_{A_2) C} K^{B_2\ldots B_{2n} )C}+b_n\,
\d_{A_1}{}^{(B_1} \d_{A_2}{}^{B_2}\del_{C D}K^{B_3\ldots B_{2n})CD}\ ,
\la{2.9}
\ee
where
\be
a_n=\frac{4n}{t+2n}\qquad b_n=-\frac{2n(2n-1)}{(t+2n)(t+2n-1)}\ ,
\la{2.10}
\ee
and where $t=2-M$ is the supertrace, and $N=2M$.
\subsubsection*{$D=4$}
\begin{align}
\del_{AA'} K^{B_1\ldots B_n,B'_1\ldots B'_n}&= a_n( \d_A{}^{(B_1}\del_{CA'} K^{B_2\ldots B_n)C,B'_1\ldots B'_n})+
a'_n(\d_{A'}{}^{(B'_1}\del_{AC'} K^{B_1\ldots B_n,B'_2\ldots B'_n)C'})\nn\w1 
&+b_n\, \d_A{}^{(B_1}\d_{A'}{}^{(B'_1}\del_{CC'} K^{B_2\ldots B_n) C,B'_2\ldots B'_n)C'}\ ,
\label{2.11}
\end{align}
where
\be
a_n=\frac{1}{t_n} \qquad a'_n=\frac{1}{t'_n} \qquad b_n=-\frac{1}{t_n t'_n}\ ,
\la{2.12}
\ee
with 
\be
t_n=\frac{n-1+t}{n}\qquad t'_n=\frac{n-1+t'}{n}\ .
\la{2.13}
\ee 
Here we have included the case of $N$ odd, $N=2M+1$. In this case the relevant analytic superspaces are the spaces of $(2|M)$-planes in $\bbC^{4|2M+1}$ and $t=2-M$ while $t'=2-(M+1)$, whereas for $N=2M$, even, $t=t'=2-M$. In the odd case there are actually two dual superspaces corresponding to flipping the number of odd directions, \ie $M\leftrightarrow (M+1)$. 
\subsubsection*{$D=6$}
\begin{align}
\del_{A_1 A_2} K^{B_1 B_2,C_1, C_2,\ldots} &=(a_n\, \d_{[A_1}{}^{[B_1} \del_{A_2] D} K^{B_2] D,C_1 C_2,\ldots} +
(n-1)\  {\rm terms})\nn\w1
&+ b_n\,(\d_{[A_1}{}^{[B_1}\d_{A_2]}{}^{B_2]}\del\cdot K^{C_1C_2,\ldots}+ {\rm cyclic})\nn\w1
&-\frac{6b_n}{n+1}(\sum \d_{[A_1}{}^{[B_1}\d_{A_2]}{}^{B_2} \del\cdot K^{C_1 C_2],D_1D_2,\ldots})\ ,
\la{2.14}
\end{align}
where in the second line the cyclic sum is over the $n$ pairs, and where the sum in the third line is over all distinct pairs of pairs, \ie $\half n(n-1)$ terms altogether. In the expression on the third line for each selected pair of pairs there is total graded antisymmetrisation. The coefficients are given by
\begin{align}
a_n&=\frac{4}{t+n-3}\nn\w1
b_n&=\frac{-(n+1)}{(t+n-2)(t+n-3)}\ .
\label{2.15}
\end{align}

A rank $n$ SCKT $K$ can be contracted with a rank $(n+1)$ symmetric conserved current to give a covector current which is conserved by virtue of the constraints that the SCKTs satisfy as well as the conservation condition on the current. 
\subsection{Super-Laplacians and minimal multipets}
The digression on currents, aside from its intrinsic interest, immediately suggests an appropriate definition of a super-Laplacian: simply replace $J$ in \eq{2.7} by a second derivative. Thus we have
\begin{align}
D=3&:\, \D_{AB,CD}= \del_{[A[C} \del_{B]D]} \ ,\nn\w1
D=4&:\D_{AB,A'B'}=\del_{[A[A'} \del_{B]B']} \ \ ,\nn\w1
D=6&: \ \, \D_{ABCD}=\del_{[AB} \del_{CD]}\ .
\label{2.16}
\end{align}
These operators all have the standard Laplacian as their highest-dimensional component, \ie the component which has only spacetime indices. 

We now briefly discuss the minimal multiplets in each case. These are single-component superfields $W$ that are annihilated by the super-Laplacians introduced above.
\subsubsection{$D=3$}
This is the simplest case. The physical fields are either scalars or spinors and occur at the zeroth and first orders in $W$ in an expansion in the odd coordinates. The internal coordinates are $y^{ab}=-y^{ba},\, a,b=1\ldots M$, where $N=2M$, and the internal Laplacian, $\D_{ab,cd}=\del_{(a(c}\del_{b)d)}$, is symmetric on $ab$ and $cd$. Expanding $W$ in powers of $y$ and imposing $\D_{ab,cd}W=0$ we find
\be
W=\sum_{n=0}^M y^n W_{(2n)}\ ,
\la{2.17}
\ee
where $W_{(2n)}$ is totally antisymmetric on its $2n$ internal indices. Differentiating this with $\del_{\a a}$ we get a series of terms with one extra internal index. But then we have
\be
\D_{ab,c\d} W\sim \del_{(a|c|}\del_{b)\d}W=0
\la{2.18}
\ee
and this implies that in each term in the $y$-expansion all of the indices have to be totally antisymmetric,
\be
\del_{\a a}W= \sum_{n=0}^{(M-1)}W_{\a a(2n)}\ ,
\la{2.19}
\ee
where $W_{\a a(2n)}=W_{\a [a_1\ldots a_{2n+1}]}$. At the next level in odd coordinates there are two components of the Laplacian
\begin{align}
\D_{ab,\c\d}&\sim \del_{(a\c}\del_{b)\d}\ ,\nn\w1
\D_{a\b,c\d}&\sim \del_{ac}\del_{\b\d} +\del_{\b c}\del_{\d a}\ .
\label{2.20}
\end{align}
Now $\del_{a\c}\del_{b\d}W$ is antisymmetric under the interchange of the pairs of indices on the derivatives and so can be either symmetric on spinor indices and antisymmetric on the internal indices or vice versa. Demanding that $\D_{ab,\c\d}W=0$ implies that we are left with the term that is symmetric on the spinor indices so that $\D_{a\b,c\d}W=0$ implies that this expression is given by a spacetime derivative acting on lower terms in the odd-variable expansion.
 This shows that the independent spacetime fields are the scalars and spin one-half fermions as claimed.
At the next level
\be
\D_{\a\b,\c d}W=0 \Rightarrow \del_\c{}^\d \del_{\d d}W=0\ ,
\la{2.21}
\ee
so that the fermion fields obey the Dirac equation. Finally the top component of the super-Laplacian is simply the spacetime Laplacian so that all of the spacetime fields in the multiplet obey the ordinary Laplace equation.

It is not difficult to see that the scalars combine into a Weyl spinor of $Spin(N)$ while the  fermions combine into the other (opposite chirality) Weyl spinor. 
\subsubsection{$D=4$}
In $D=4$ the minimal multiplets are free massless multiplets with maximal helicity $N/4$. The component field-strengths divide into two sets having completely symmetrised primed or unprimed indices. The lowest-order component of the super-Laplacian is $\D_{ab,a'b'}=\del_{(a(a'}\del_{b)b')}$ so that imposing that the superfield $W$ be annihilated by this implies that it has a $y$-expansion of the form
\be
W=\sum_{n=0}^M y^n W_{(n,n)}\ ,
\la{2.22}
\ee
where $W_{(n,n)}$ is totally antisymmetric on $n$ unprimed and $n$ primed internal even indices, \ie $W_{(n,n)}=W_{[a_1\ldots a_n],[a'_1\ldots a'_n]}$.

If we apply odd derivatives to this expression \la{2.22} and demand that the relevant components  of the super-Laplacian acting on the resulting fields vanish, we find, as in the $D=3$ case, that terms with only primed or unprimed spinor indices have to be completely symmetrised in them, while terms with mixed primed and unprimed spinor indices involve spacetime derivatives of lower-order terms. Moreover, the $y$-expansions of the pure terms involve components that are anti-symmetrised with respect to both sets of primed and unprimed internal indices.  Thus at order $p$ in the odd variables  there are two fields $W_p$ and $W'_p$ having $p$ totally symmetrised unprimed spinor indices and $p$ totally symmetrised primed spinor indices respectively. These fields also carry $p$ totally antisymmetrised internal primed (unprimed) indices respectively, as required by the super Laplace equation. Both of these fields have to be expanded in the $y$-variables, so that, for example,
\be
W_p\sim \sum_{n=0}^{(M-p)} y^n W_{(n,p+n)}
\la{2.23}
\ee
where each component $W_{(n,p+n)}$ has $n$ antisymmetrised unprimed internal indices and $(n+p)$ antisymmetrised primed internal indices, as well as $p$ symmetrised unprimed spinor indices. Clearly the top components will have $M$ antisymmetrised spinor indices and will therefore only have one independent internal symmetry component. For $p\geq2$ these field, evaluated at $\xi=\xi'=0$ are spacetime field-strengths. The other components of the super-Laplacian then imply that the fields with at least one spinor index obey spacetime equations of the form
\be
\del^{\a_1}{}_{\bdt} W_{\a_1\ldots \a_p}=0\ .
\la{2.24}
\ee
These are the equations of motion for the spin-one-half fields and the equations of motion combined with the Bianchi identities for the higher-spin fields.

Some examples are the hypermultipet for $N=2$, the vector multiplet for $N=4$ and the maximal supergravity multiplet for $N=8$. For $N=4$ we have
\begin{align}
W&=W_0 + y^{aa'} W_{aa'} + y^{aa'}y^{bb'} W_{ab,a'b'}\ ,\nn\w1
\del_{\a a'}W&=W_{\a a'}+ y^{bb'} W_{\a b,a'b'}\ ,\nn\w1
\del_{a \a'}W&=W_{\a' a'}+ y^{bb'} W_{\a ab,a'}\ ,\nn\w1
\del_{\a a'}\del_{\b b'}W&=W_{\a\b a'b'}\ ,\nn\w1
\del_{\a' a}\del_{\b' b}W&=W_{\a'\b' ab}\ ,
\label{2.25}
\end{align}
where all of the spinor indices are symmetrised and all of the internal indices are antisymmetrised. A similar analysis of the $N=8$ case leads to the expected spectrum, $70+56+28+8+1$, of component fields with spins $0,1/2,1/3/2$ and $2$, respectively, subject to their equations of motion and Bianchi identities.
\subsubsection{$D=6$}
In the $D=6$ case the internal coordinates $y^{ab}$ are symmetric in $a,b=1\ldots N$ (not $N/2$), and the lowest component of the super-Laplacian is $\D_{abcd}=\del_{(ab}\del_{cd)}$. Demanding that this be zero when applied to a superfield $W$ then gives a $y$-expansion of the form
\be
W=\sum_{n=0}^N y^n W_{(n,n)}\ ,
\la{2.26}
\ee
where the coefficients $W_{(n,n)}$ correspond to fields with the symmetries of Young tableaux with two columns each with $n$ boxes. At the next level the super-Laplacian is $\del_{(ab}\del_{c)\d}$, so that the fields with one spinor index are of the form $W_{(n+1,n)}$ corresponding to the Young tableaux with $(n+1)$ boxes in the first column and $n$ in the second. The pattern continues at higher orders. The new independent fields at level $p$ have $p$ symmetrised spinor indices and have $y$-expansions with coefficient  fields of the form $W_{(n+p,n)}$ corresponding to two-column Young tableaux with $(n+p)$ boxes in the first column and $n$ in the second. From the higher-order components of the super-Laplacian it follows that the spin-half fermions $W_\a$ obey the Dirac equation
\be
\del^{\a\b} W_\b=0; \qquad \del^{\a\b}:=\half \ve^{\a\b\c\d} \del_{\c\d}\ ,
\la{2.27}
\ee
while at level $p$ we have
\be
\del^{\a\c} W_{\c \b_1 \ldots \b_{(p-1)}} - \frac{p-1}{p+1}\d^\a_{(\b_1} \del^{\c\d} W_{\c \b_1 \ldots \b_{(p-2)})\c\d}=0\ ,
\la{2.28}
\ee
an equation which combines the Bianchi identities and equations of motion for the higher-spin fields.
The simplest example, $N=1$, is again the hypermultiplet, while for $N=2$ we have the $(2,0)$ tensor multiplet. The components of this multiplet are
\begin{align}
W&=W_0+y^{ab} W_{ab}+y^{ab} y^{cd} W_{ab,cd}\nn\w1
\del_{\a a}W&= W_{\a a} + y^{bc} W_{\a a,bc}\nn\w1
\del_{(\a a}\del_{\b) b}W&=W_{\a\b ab}\ .
\label{2.29}
\end{align}
The components of this multiplet are therefore $1+3+1=5$ scalars $2+2=4$ spin one-half fields and a single symmetric bi-spinor,
$W_{\a\b}$, equivalent to a self-dual three-form field-strength tensor in spacetime. 

As a final example we consider $N=4$. This is an alternative maximal (linearised) supergravity multiplet proposed in \cite{Hull:2000zn} in which the gravitational degrees of freedom are represented by a symmetric four-spinor equivalent to a spacetime tensor with two sets of self-dual three-form indices. This multiplet has $42$ scalars, $48$ spin one-half fields, $27$ self-dual three-forms, $8$ spin three-halves fields represented by symmetric three-spinor field strengths and a single four-spinor, all fields being subject to \eq{2.27} or \eq{2.28} as well as the spacetime Laplace equation.
\subsubsection{Comment}
In the above examples we have shown how single-component fields on analytic superspaces satisfying the super-Laplace equations describe the minimal supermultiplets for each $D$ and $N$ ( $N$ even in $D=3,4$). The discussion given is entirely local even in the internal coordinates. In practice we normally want the fields to be defined on the whole of the internal cosets which are compact complex manifolds. This means that they automatically have short expansions in the internal coordinates by analyticity. For example, for $D=4, N=4$ the internal coset is $S(U(2)\xz U(2))\bsh SU(4)$, so that a scalar field $W$ with $U(1)$ charge 2 with respect to the central node of the $\gs\gu(4)$ Dynkin diagram will automatically describe an on-shell Maxwell multiplet. Some of the other examples discussed here have also been discussed from this point of view in the literature, particularly the hypermultiplet \cite{Galperin:1984av} and 16 supersymmetry (half-maximal) multiplets in $D=3,4,6$ \cite{Howe:1998jw}. In the local description used above, the weight of a singlet superfield $W$ is uniquely fixed by the requirement that the super-Laplacian acting on $W$ be covariant under superconformal transformations.

An alternative approach to the coordinate description we have employed here would to be to use equivariant methods, \ie to work on the relevant superconformal group but specify a field's properties with respect to the relevant isotropy group. This method was employed to give a discussion of semi-shortening conditions for superconformal representations in \cite{Ju:2013mkc}.

\section{Symmetries}

In this section we discuss the construction of symmetries of super-Laplacians in analytic superspaces in terms of super-Killing tensors. We shall show that we can systematically construct all of the terms of a symmetry $\cD_K$ from its leading term in terms of an $n$th rank SCKT $K$ and its derivatives. For all cases we can set
\be
\cD_K=\sum_{m=0}^{m=n} \a_m (\del^m\cdot K) \del^p\ ,
\la{3.0.1}
\ee
where $K$ is an $n$th rank SCKT, $n=m+p$ and the $\a_m$ are constants, with $\a_0$ taken to be 1. For all cases $K$ has $2n$ indices and each derivative has two indices. The notation indicates that $2m$ of the indices on $K$ are contracted with those of the $m$ derivatives which are applied to it, while the remaining $2p$ indices are contracted with those on the final $p$ derivatives. The super-Laplacian consists of two derivatives projected onto the appropriate representation of the isotropy group as in \eq{2.16}. The leading term in \eq{3.0.1} has the form $K\del^n$, from which it apparent that $K$ must have the index structure of a SCKT since otherwise the derivatives would include at least one super-Laplacian. In fact, for all three cases, the product of two derivatives falls into two representations of which one is the super-Laplacian and can therefore be ignored in $\cD_K$.

Applying the super-Laplacian to \eq{3.0.1} we find a series of constraints of the form
\be
\a_m \del\del (\del^m\cdot K) \del^p + 2\a_{m+1} \del (\del^{m+1}\cdot K) \del\del^{p-1}\sim 0\ ,
\la{3.0.2}
\ee
where $m=-1,\ldots n$, with $\a_{-1}=\a_{n+1}=0$, and where the appropriate projection on the two extra derivatives is understood. Terms involving the super-Laplacian to the right of $K$ are to be disregarded, ({\sl cf.} \eq{1.1}). The first term in the series \eq{3.0.2}, \ie $m=-1$ involves terms with $\del\del^n$ and has only one contribution, from the $\a_0$ term. It is easy to see that this equation implies that $K$ must be a SCKT. The task is then to show that all the remaining terms can be constructed sequentially. This indeed turns out to be the case, and there are unique solutions for all cases in $D=3,4,6$. More precisely, what this shows is that given an $n$-th rank symmetry, its leading term is given by an $n$th-rank SCKT $K$ contracted into $n$ derivatives, and that there are canonical lower-order terms given in terms of the derivatives of  $K$ according to \eq{3.0.1}. There could be other lower-order terms involving lower-rank SCKTs. 

We shall discuss the proof of the above assertion below, although separate derivations are required for each dimension $D=3,4,6$. 
\subsection*{$D=3$}
We begin with the simplest example, $n=1$ in $D=3$. Such a symmetry operator has the form
\be
\cD_K=K^{AB}\del_{AB} + K_0\ .
\la{3.1}
\ee
where $K_0$ is a function. When we apply the super-Laplacian to this object we shall get differential operators with zero, one or two derivatives. Some of the latter can be discarded because they are themselves proportional to the super-Laplacian. Let us consider the terms with two derivatives; the $K_0$ term can be ignored since it is the Laplacian, so we require
\be
\del_{[A[C} K^{EF}\del_{B]D]}\del_{EF}   \sim 0\ ,
\la{3.2}
\ee
where $\sim$ indicates that terms proportional to the Laplacian can be dropped, and where the graded antisymmetrisation is over the pairs $AB$ and $CD$ separately. It is immediately apparent that the non-trace part of $\del_{AB} K^{CD}$ must be zero because otherwise this equation cannot be satisfied. This means that $\del_{AB} K^{CD}$ must involve unit tensors, and by consistency $K$ must be a superconformal Killing vector $K$ satisfying \eq{2.9} for the case $n=1$. Note that indecomposability is not a problem here, the constraint is simply that the full $\del K$ tensor is given by sub-representations. For the terms with one derivative we must have
\be
\del_{[A[C} \del_{B]D]} K^{EF}\del_{EF} + 2\del_{[A[C} K_0 \del_{B]D]}=0\ . 
\la{3.3}
\ee
It is not difficult to show that
\be
\del_{AB}\del_{CD} K^{EF}=\frac{2}{(t+1)} \d_A{}^E\d_C{}^F\del_{BD}\, \del\cdot K
\la{3.4}
\ee
where symmetrisation over the pairs $AB,CD$ and $EF$ is understood, and where $t$ is the supertrace. Substituting this result back in \eq{3.3} we see that this equation will be satisfied if
\be
\cD_K=K^{AB}\del_{AB} + \frac{1}{2(t+1)}\, \del\cdot K\ .
\la{3.5}
\ee

For a rank 2 SCKT $K$ we find
\be
\cD_K=K^{ABCD} \del_{AB}\del_{CD} +\frac{3}{(t+3)} \del_{AB} K^{ABCD} \del_{CD} + \frac{3}{4(t+2)(t+3)} \del_{AB}\del_{CD} K^{ABCD}\ ,
\la{3.7}
\ee
where the coefficients are determined by the requirement that $\cD_K$ be a symmetry of the super-Laplacian. 

We can extend this argument to the general case. It is clear that the leading term for an $n$th rank differential operator must be
\be
K^{A_1\ldots A_{2n}} \del^n_{A_1\ldots A_{2n}}
\la{3.6}
\ee
where $K$ is totally graded symmetric and where the second factor denotes the totally graded symmetric product of $n$ derivatives. This must be the case because any non-symmetrisation would immediately lead to a super-Laplacian amongst the derivatives, as mentioned previously. This is perhaps easier to see if one uses the fact that the product of two derivatives has just two representations, the totally symmetric part and the super-Laplacian, the latter corresponding to a $2\xz 2$ square Young tableau. 

It is straightforward to compute the differential symmetry operators $\cD_K$ corresponding to higher-order SCKTs. For an $n$th-order SCKT put
\be
(\del^m\cdot K)^{F_1\ldots F_{2p}}=\del_{E_1E_2}\ldots \del_{E_{2m-1}E_{2m}} K^{E_1\ldots E_{2m} F_1\ldots F_{2p}}\ ,
\la{3.7.0.1}
\ee
where $K$ is totally symmetric on all $2n$ indices. We wish to compare two terms
\begin{align}
T_1&\sim \a_m \del\del (\del^m\cdot K) \del^p\ \ \  {\rm and}\nn\w1
T_2&\sim 2\a_{m+1} \del (\del^{m+1}\cdot K) \del\del^p\ ,
\label{3.7.0.2}
\end{align}
where the two unindexed derivatives are to be projected onto the super-Laplacian representation. Now consider
\be
\del_{AC}\del_{BD}(\del^m\cdot K)^{F_1\ldots F_{2p}}=A_{AB,CD}{}^{F_1\ldots F_{2p}}+B_{AB,CD}{}^{F_1\ldots F_{2p}}\ ,
\la{3.7.1}
\ee
where $A$ is antisymmetric on both pairs of indices, $AB$ and $CD$, and symmetric on the interchange of the pairs while $B$ is symmetric on both pairs and symmetric on their interchange. It follows from the definitions of SCKTs in analytic superspace (for any $D$) that each derivative which is applied to a SCKT will give rise to a new unit tensor (delta). We can therefore write
\be
A_{AB,CD}{}^{F_1\ldots F_{2p}}=4\d_{[A}^{(F_1}\d_{[C}^{F_2} A_{B]D]}{}^{F_3\ldots F_{2p})}\ ,
\la{3.7.2}
\ee
where $A_{BD}$ is symmetric. Note that, on the $AB,CD$ indices, $A$ in \eq{3.7.1} is in the irreducible representation corresponding to the Riemann tensor, with Young tableau (YT) (2,2), where in text we use the notation $(p,q,r,\ldots)$ to denote a YT with $p$ boxes in the first row, $q$ in the second, and so on. $B$ is reducible; it can contain the (2,2) representation as well as the totally symmetric one with YT (4). The Laplacian corresponds to the (2,2) representation so that the $B$ term will drop out of the $\a_m$ term  in $T_1$ (\eq{3.7.0.2}) due to the projection onto the super-Laplacian representation on the indices $AB,CD$. 

The $\a_{m+1}$ term will involve
\begin{align}
\del_{AC}(\del^{(m+1)}\cdot K)^{F_3\ldots F_{2p}}&=\del_{AC}\del_{F_1 F_2} (\del^m\cdot K)^{F_1 F_2 F_3\ldots F_{2p}}\nn\w1
&=4\d_{[A}^{(F_1}\d_{[C}^{F_2} A_{F_1]F_2]}{}^{F_3\ldots F_{2p})} + B_{AC}{}^{F_3\ldots F_{2p}}\ ,
\label{3.7.3}
\end{align}
where $B_{AC}{}^{F_3\ldots F_{2p}}=B_{AB,CD}{}^{BD F_3\ldots F_{2p}}$ is symmetric on $AC$. The $A$ term in \eq{3.7.3} is equal to $f(t) A_{AC}{}^{F_3\ldots F_{2p}}$ where 
\be
f(t)=\frac{(t+2p-2)(t+2p-3)}{2p(2p-1)}\ ,
\la{3.7.4}
\ee
and where we have dropped terms involving $\d_A^F$ or $\d_C^F$. These will not contribute when we evaluate the term $\del(\del^{(m+1)}\cdot K)\del \del^{(p-1)}$, because they would give rise to Laplacian combinations of two derivatives to the right of $K$.

To derive a relation between $A$ and $B$ we consider $\del_{AC}(\del^{(m+1)}\cdot K)^{F_3\ldots F_{2p}}$, let $\del_{AC}$ act on $K$ and use the equations \eq{2.9} for a SCKT of rank $n$. After a little algebra one finds
\be
\k\,\del_{AC}(\del^{(m+1)}\cdot K)^{F_3\ldots F_{2p}}=\l\,\del_{AB}\del_{CD}(\del^{m}\cdot K)^{BD F_3\ldots F_{2p}}\ ,
\la{3.7.5}
\ee
where 
\begin{align}
\k&=1-\frac{(m+1)b_n}{n(2n-1)}\nn\w1
\l&=\frac{(m+1)a_n}{n}+ \frac{2m(m+1) b_n}{n(2n-1)}\ ,
\la{3.7.6}
\end{align}
and where we have again dropped irrelevant delta-terms.

The LHS of \eq{3.7.5} is $fA+B$ while the RHS is $B-fA$, (both with free $AC$ lower indices), while $T_1\sim 4A$ and $T_2\sim 2(fA+B)$. We therefore find that the ratio of two successive $\a$ coefficients in \eq{3.0.2} is given by
\be
\frac{\a_{m+1}}{\a_m}=\frac{p(2p-1)}{2(m+1)(t+2n-(m+1))}\ .
\ee

This agrees with the formul{\ae}  \eq{3.7} and \eq{3.9} for the cases $n=1$ and $2$ respectively.

\subsection*{$D=4$}
In $D=4$ a symmetry corresponding to a SCKV $K$ has the form
\be
\cD_K= K^{AA'}\del_{AA'} + K_0
\la{3.8}
\ee
For $\cD_K$ to be a symmetry it is not diffcult to show that $K_0\propto \del\cdot K$; explicitly
\be
\cD_K= K^{AA'}\del_{AA'} + \frac{1}{2t} \del\cdot K\ .
\la{3.9}
\ee
using arguments similar to those for the $D=3$ case we find that a second-rank symmetry has the form
\be
\cD_K=K^{ABA'B'}\del_{AA'} \del_{BB'} +\frac{2}{(t+1)} \del_{AA'} K^{ABA'B'} \del_{BB'} +\frac{1}{(t+1)(2t+1)} \del_{AA'} \del_{BB'} K^{ABA'B'}\ .
\la{3.10}
\ee

For the general case of an $n$th rank symmetry put

\be
(\del^m\cdot K)^{E_1\ldots E_p E'_1\ldots E'_p}=\del_{C_1 C'_1}\ldots \del_{C_m C'_m} K^{C_1\ldots C_m E_1\ldots E_p C'_1\ldots C'_m E'_1\ldots E'_p}\ ,
\la{3.10.1}
\ee
where the $E$-indices are symmetrised. Consider
\be
\del_{AA'}\del_{BB'} (\del^m\cdot K)^{E_1\ldots E_p E'_1\ldots E'_p}=4\d_{[A}^{(E_1}\d_{[A'}^{(E'_1}A_{B]B']}{}^{E_2\ldots E_p)E'_2\ldots E'_p)}+B_{ABA'B'}{}^{E_1\ldots E_p E'_1\ldots E'_p}\ ,
\la{3.10.2}
\ee
where $A$ is antisymmetric on $AB$ and $A'B'$ and $B$ is symmetric on both pairs. The $\d$ structure of $A$ is required for similar reasons to the $D=3$ case.

We want to compare two terms,  $T_1$ and $T_2$ say. The first is given by
\be
T_1=\del_{[A[A'}\del_{B]B']}(\del^m\cdot K)^{E_1\ldots E_p E'_1\ldots E'_p} {\del^p}_{E_1\ldots E_p E'_1\ldots E'_p}
=4 A_{[A[A'}{}^{E_2\ldots E_p E'_2 \ldots E'_p}{\del^p}_{B] E_2\ldots E_p B' ]E'_2\ldots E'_p}
\la{3.10.2}
\ee
where $\del^p$ denotes $p$  derivatives, symmetrised on both primed and unprimed indices. Note that the trace terms in $A$ do not contribute because they would lead two external indices $A,B$ etc on the derivatives to the right and these would give rise to Laplacians. The second term is

\begin{align}
\half T_2&=\del_{[A[A'}(\del^{m+1}\cdot K)^{E_2\ldots E_p E'_2\ldots E'_p} \del_{B]B']} \del^{p-1}_{E_2\ldots E_p E'_2\ldots E'_p}\nn\w1
&=\left(4\d_{[A}^{(C}\d_{[A'}^{(C'}A_{C]C']}{}^{E_2\ldots E_p)E'_2\ldots E'_p)}+B_{AA'}{}^{E_2\ldots E_p E'_2\ldots E'_p}\right)\del_{B]B'] }
\del^{p-1}_{E_2\ldots E_p E'_2\ldots E'_p}\ .
\la{3.10.3}
\end{align}
Here $B_{AA'}$ is the trace of the original $B_{AB,A'B'}$ above over one set of primed and one set of unprimed indices. Note that trace terms in $A$ and $B$ do not contribute to $T_2$ because of antisymmetry. The first term in the last line can be evaluated straightforwardly and one finds
\be
\half T_2=(f(t) A_{[A[A'}{}^{E_2\ldots E_p E'_2\ldots E'_p}+B_{BB'}{}^{E_2\ldots E_p E'_2\ldots E'_p})\del_{B]B']} \del^{p-1}_{E_2\ldots E_p E'_2\ldots E'_p}\ ,
\la{3.10.4}
\ee
with the antisymmetry indicated. Here
\be
f(t)=\frac{1}{p^2}(t+p-2)^2\ .
\la{3.10.5}
\ee
Note that in going from (12) to (13) we have dropped terms from the $A$ term which contain $\d_A^E$ factors because they do not contribute to $T_2$.

To derive a relation between $A$ and $B$ we consider $\del_{AA'}(\del^{m+1}\cdot K)^{E_2\ldots E_p E'_2\ldots E'_p}$, with the $E$s symmetrised, let $\del_{AA'}$ act on $K$ and use the SCKT equations. There are terms that involve $\d_A^E$s that can be ignored as they lead to Laplacian contributions to $T_1$ or $T_2$. One finds
\be
\k \del_{AA'}(\del^{m+1}\cdot K)^{E_2\ldots E_p E'_2\ldots E'_p}=\l \del_{AA'}(\del^{m+1}\cdot K)^{E_2\ldots E_p E'_2\ldots E'_p}\ ,
\la{3.10.6}
\ee
where
\begin{align}
\k&=\left(1-\frac{(m+1)}{n}(2a_n+(m+1)\frac{b_n}{n})\right)\nn\w1
\l&=-4\frac{(m+1)}{n^2}(na_n+\frac{mb_n}{2})\ .
\label{3.10.7}
\end{align}

This can be rewritten as
\be
\k(f(t) A + B)=-\l f(t)B
\la{3.10.8}
\ee
where all the indices on the traced $A$ and $B$ have been supressed. Now both $\k$ and $\l$ can be expressed as fractions with $(t+n-1)^2$ in the denominators, coming from $b_n$, and these will cancel in \eq{3.10.8}. So we can rewrite this equation in terms of $\bar\k$ and $\bar\l$, say, where the common denominator factors have been removed. Now actually we want to compare $T_1\sim 4A$ with $T_2\sim 2(fA+B)$. Simple computations give
\be
\bar\k=p^2 f;\qquad \bar\l=2(m+1)(2t+m+2p-2)
\la{3.10.9}
\ee
Using \eq{3.10.8} we then find
\be
T_2=2(fA+B)=-2\frac{\bar\l}{\bar\k}fA=-2\frac{\bar\l}{p^2} A=-\half
\frac{\bar\l}{p^2} T_1
\ee
From which we get
\be
\frac{T_2}{T_1}=\frac{1}{p^2}(m+1)(2t+m+2p-2)\ .
\la{3.10.10}
\ee
In terms of the coefficients $\a_m$  introduced earlier we have
\be
\a_{m+1}=\frac{p^2}{(m+1)(2t+m+2p-2)}\a_m\ ,
\la{3.10.11}
\ee
which agrees with the $n=1,2$ results given explicitly above.
\subsection*{$D=6$}
In $D=6$ similar arguments show that a symmetry corresponding to a SCKV $K$ has the form
\be
\cD_K= K^{AB}\del_{AB} + \frac{1}{(t-1)} \del\cdot K\ .
\la{3.11}
\ee
while a second-rank symmetry has the form
\be
\cD_K=K^{AB,CD} \del_{AB}\del_{CD} +\frac{3}{t} \del_{AB} K^{ABCD} \del_{CD} + \frac{3}{t(2t-1)} \del_{AB}\del_{CD} K^{ABCD}\ .
\la{3.12}
\ee

In $D=6$ the coordinates are antisymmetric $X^{AB}=-X^{BA}$ so that the product of two derivatives falls into two irreducible representations of the isotropy group, totally antisymmetric with  YT (1,1,1,1), and  the Riemann tensor, (2,2). An $n$th rank SCKT has $2n$ indices, antisymmetric on $n$ pairs, totally symmetric with respect to the pairs and such that anti-symmetrisation over any three indies gives 0. The Young tableau is $\overbrace{\tiny\yng(8,8)}^{n}$ or just $(n,n)$. We set
\be
\del_{AB}\del_{CD} (\del^m\cdot K)^{E_1 F_1,\ldots E_p,F_p}=A_{ABCD}{}^{E_1 F_1,\ldots E_p,F_p}+ B_{AB,CD}{}^{E_1 F_1,\ldots E_p,F_p}\ ,
\la{3.12.1}
\ee
where $A$ is totally antisymmetric on $ABCD$ while $B$ is antisymmetric on both pairs $AB$ and $CD$ and symmetric under the interchange of the pairs. So in this case, both $A$ and $B$ are irreducible with respect to their lower indices. We can write
\be
A_{ABCD}{}^{E_1F_1,E_2 F_2,\ldots E_p,F_p}=\d_{[AB}^{E_1 F_1} A_{CD]}{}^{E_2 F_2,\ldots E_p,F_p}\ ,
\la{3.12.2}
\ee
where the two-index delta denotes the anti-symmetrised product of two deltas.

The first term $T_1$ is then given by
\be
T_1=A_{[AB}{}^{E_2 F_2,\ldots E_p,F_p}\del_{CD]} \del^{p-1}_{E_2 F_2,\ldots E_p,F_p}\ ,
\la{3.12.3}
\ee
where the final $(p-1)$ derivatives are in the $(p-1,p-1)$ representation. The second term $T_2$ is given by
\be
\half T_2=\del_{[AB}(\del^{m+1}\cdot K)^{E_2 F_2,\ldots E_p,F_p}\del_{CD]}\del^{p-1}_{E_2 F_2,\ldots E_p,F_p}\ .
\la{3.12.4}
\ee
Using (4.1) and (4.2) we can see that the RHS of this, omitting the derivatives to the right,  is
\be
f(t) A_{AB}{}^{E_2 F_2,\ldots E_p,F_p} + B_{AB}{}^{E_2 F_2,\ldots E_p,F_p}\ ,
\la{3.12.5}
\ee
where 
\be
B_{AB}{}^{E_2 F_2,\ldots E_p,F_p}:=B_{AB,E_1F_1}{}^{E_1 F_1,\ldots E_p,F_p}\ ,
\la{3.12.6}
\ee
and where
\be
\d_{[AB}^{E_1 F_1} A_{E_1 F_1]}{}^{E_2 F_2,\ldots E_p,F_p}=f(t) A_{AB}{}^{E_2 F_2,\ldots E_p,F_p}\ .
\la{3.12.7}
\ee
Schematically,
\be
T_1\sim A:\qquad T_2\sim 2(fA + B)\ .
\la{3.12.8}
\ee
To calculate $f$ and the relation between $A$ and $B$ one has to make repeated use of the projection of $X^{C_1D_1} Y^{C_2 D_2,\ldots C_n D_n}$, where $X$ is skew and $Y$ in the representation with Young tableau $(n-1,n-1)$ onto the irreducible representation with Young tableau $(n,n)$. This is given by
\begin{align}
Z^{C_1 D_1,\ldots C_n D_n}&=\frac{1}{n}\left( X^{C_1 D_1}Y^{C_2 D_2,\ldots C_n D_n} + {\rm cyclic\ over\ pairs}\right)\nn\w1
&-\frac{6}{n(n+1)}\sum \left( X^{[C_1 D_1}Y^{C_2 D_2],\ldots C_n D_n} +\ldots\right)\ ,
\la{3.12.9}
\end{align}
where the sum in the second line is over all $\half n(n-1)$ distinct selected pairs of pairs.

A calculation yields
\be
f(t)=\frac{1}{6p(p+1)}\left((t+p-3)(t+p-4)\right)\ ,
\la{3.12.10}
\ee
and, starting from $\del_{AB}(\del^{m+1}\cdot K)^{E_2 F_2,\ldots E_p,F_p}$, we find the relation
\be
\k\del_{AB}(\del^{m+1}\cdot K)^{E_2 F_2,\ldots E_p,F_p}=\l\del_{AG}\del_{BH}(\del^m\cdot K)^{GH,E_2 F_2,\ldots E_p,F_p}\ ,
\la{3.12.11}
\ee
with
\begin{align}
\k&=(1-\frac{2(m+1)b_n}{(n+1)})\nn\w1
\l&=(m+1)(a_n+\frac{2m b_n}{(n+1)})\ .
\label{3.12.12}
\end{align}
One can show that
\be
\del_{AG}\del_{BH}(\del^m\cdot K)^{GH,E_2 F_2,\ldots E_p,F_p}\sim - fA + \half B
\la{3.12.13}
\ee
so that, using (4.8) one gets
\be
\frac{T_1}{T_2}=\frac{\l-2\k}{6\l f}\ .
\la{3.12.14}
\ee
Using the values for $a_n$ and $b_n$ given in \eq{2.15} one finds
\be
\frac{\a_{m+1}}{\a_m}=-\frac{T_1}{T_2}=\frac{p(p+1)}{(m+1)(2t+2n-m-4)}\ .
\ee
This agrees with the results for $n=1,2$ given above.

\section{Products of symmetries}
\subsection{$D=3$}
Consider two SCKVs in $D=3$. The corresponding symmetries $\cD_K$ and $\cD_L$ have the forms given in \re{3.1}. Multiplying these together gives terms with zero, one and two derivatives. Explicitly,
\begin{align}
\cD_K \cD_L&= K^{AB} L^{CD} \del_{AB}\del_{CD} + (K^{CD}\del_{CD} L^{AB} + K_0 L^{AB}+K^{AB} L_0)\del_{AB}\nn\w1
&\ \ +(K^{AB} \del_{AB} L_0 + K_0 L_0)\ .
\label{4.1}
\end{align}
In the two-derivative term we can discard the term proportional to the super-Laplacian leaving $K^{(AB} L^{CD)}\del^2_{ABCD}$ which we are going to identify with the leading term in $\cD_M$, where $M=K\circledcirc L$. The terms with one derivative give
\begin{align}
&(K^{CD}\del_{CD} L^{AB} + K_0 L^{AB}+K^{AB} L_0)\del_{AB} + K_0 L^{AB}+K^{AB} L_0)\del_{AB}=\nn\w1
&\half [K,L]^{AB}\del_{AB}
 + \half\left((K^{CD}\del_{CD} L^{AB} + L^{CD}\del_{CD} K^{AB}) + (K_0 L^{AB}+K^{AB} L_0)\right)\del_{AB}
\label{4.2}
\end{align}
The first term on the right-hand-side is the one-derivative term in $\half \cD_{[K,L]}$, so the rest has to be identified with the one-derivative term in $\cD_M$ times a constant $\a$. A straightforward computation confirms this provided that $\a=3/{(t+3)}$, in agreement with \eq{3.7}.

The terms with zero derivatives should then equate to 
\be
\frac{1}{4(t+1)}\del_{AB} [K,L]^{AB} +\frac{3}{4(t+2)(t+3)}\del_{AB}\del_{CD} M^{ABCD} +K\cdot L \ ,
\la{4.3}
\ee
where the left-over term is a new scalar $K\cdot  L$. A calculation gives
\begin{align}
K\cdot L&=\frac{1}{8(t+1)} (L^{AB}\del_{AB}(\del\cdot K)+K\leftrightarrow L) +\frac{1}{4(t+1)^2 (t+2)} (\del\cdot K)(\del\cdot L)\nn\w1
&-\frac{1}{4(t+2)}\del_{AB} K^{CD}\del_{CD} L^{AB}\ .
\label{4.4}
\end{align}
This scalar should be constant in order for it to be a symmetry by itself of the super-Laplacian, and one can verify by a straightforward computation that this is indeed the case.
\subsection{$D=4$}
We can repeat the above argument for two SCKVs $K,L$  in $D=4$. We find
\be
\cD_K\cD_L=\cD_M + \half \cD_{[K,L]} + K\cdot L\ ,
\la{4.5}
\ee
where the scalar term has to be constant in order to be a symmetry. The second-rank SCKT $M$ is given by
\be
M^{ABA'B'}=(K\circledcirc L)^{ABA'B'}= K^{(A(A'}L^{B)B')}\ ,
\la{4.6}
\ee
while the scalar is found to be
\begin{align}
K\cdot L&=\frac{1}{4(2t+1)}\left( K^{AA'} \del_{AA'}\del \cdot L + L^{AA'} \del_{AA'} \del\cdot K +\frac{1}{t^2}(\del\cdot K)(\del\cdot L)\right.\nn\w1
&\ \ \qquad  \left. -2\del_{AA'} K^{BB'} \del_{BB'} L^{AA'}\right) \ .
\label{4.7}
\end{align}
\subsection{$D=6$}
For $D=6$ we find, in a similar vein, 
\be
\cD_K\cD_L=\cD_M + \half \cD_{[K,L]} + K\cdot L\ ,
\la{4.8}
\ee
where again the scalar term has to be constant in order to be a symmetry. The second-rank SCKT $M$ is given by
\be
M^{AB,CD}=(K\circledcirc L)^{AB,CD}= \frac{1}{2}\left((K^{AB}L^{CD}-K^{C[A} L^{B]D})+(K\leftrightarrow L)\right)\ ,
\la{4.9}
\ee
while the scalar is found to be
\begin{align}
K\cdot L&=\frac{1}{(2t-1)}\left( \frac{(t-2)}{2(t-1)}(K^{AB} \del_{AB} (\del \cdot L )+ L^{AB} \del_{AB}(\del \cdot K) +\frac{1}{(t-1)^2}(\del\cdot K)(\del\cdot L)\right.\nn\w1
&\ \ \qquad  \left. -\del_{AB} K^{CD} \del_{CD} L^{AB}\right) \ .
\label{4.10}
\end{align}

\section{Super-twistor space}

In this section we generalise to the supersymmetric case an argument given in \cite{Eastwood:2007} in which the Joseph ideal for the Lie algebra $\gs\gl(n)$ is discussed by considering differential operators acting linearly on $\bbC^n$. For $D=4$ the superconformal algebra acts linearly on super-twistor space, $\bbC^{4|N}$. For a traceless tensor $K_\cA{}^\cB$, which  gives an element of the super algebra $\gs\gl(4|N)$, we can define a differential operator $\cD_K$ by
\be
\cD_K= z^\cA K_\cA{}^\cB \del_{\cB}\ ,
\la{5.1}
\ee
where $z^\cA$ are standard coordinates for $\bbC^{4|N}$. Multiplying two such operators together we find
\be
\cD_K \cD_L=z^{\cB} z^\cA K_\cA{}^\cC L_\cB{}^\cD \del_\cD\del_\cC + z^\cA(KL)_\cA{}^\cB \del_\cB\ .
\la{5.2}
\ee
Clearly the coordinates and derivatives in the first term imply that the tensor constructed from $K$ and $L$ is symmetric on $\cA \cB$ and on $\cC\cD$.  This object can generically be decomposed into irreducible (\ie traceless) parts as follows
\be
K_{(\cA}{}^{(\cC} L_{\cB)}{}^{\cD)}=M_{\cA\cB}{}^{\cC\cD}+ \frac{2}{(t+2)}\d_{(\cA}{}^{(\cC} N_{\cB)}{}^{\cD)}-\frac{1}{(t+1)(t+2)}\d_{\cA\cB}{}^{\cC\cD} K\cdot L\ ,
\la{5.3}
\ee
where $M$ denotes the traceless part, $N_\cA{}^\cB=\half (KL + LK)_{\cA}{}^{\cB}$, $K\cdot L=\tr(KL)$ and the delta in the last term  indicates the product of two deltas with symmetrised indices. We can then rewrite \eq{5.2} as
\begin{align}
\cD_K \cD_L&=\cD_M +\half \cD_{[K,L]} 
+ (z^\cA N_\cA{}^\cB \del_\cB + \frac{2}{(t+2)} z^\cA z^\cB N_\cB{}^\cC \del_\cC \del_\cA)\nn\w1
&-\frac{K\cdot L}{(t+1)(t+2)}z^\cA z^\cB \del_\cB\del_\cA \ ,
\label{5.4}
\end{align}
where
\be
\cD_M=z^\cB z^\cA M_{\cA\cB}{}^{\cC\cD} \del_{\cD} \del_{\cC}\ ,
\la{5.5}
\ee
a formula that can be generalised to arbitrary traceless tensors with $n$ upper and lower symmetrised indices. The two-derivative term involving $N$ can be rewritten as
\be
\frac{2}{(t+2)}\left( z^\cB N_\cB{}^\cC \del_\cC (z^\cA \del_\cA)- z^\cB N_\cB{}^\cA \del_\cA \right)\ ,
\la{5.6}
\ee
so that the right-hand-side of \eq{5.4}, omitting the first two terms, gives
\be
\frac{t}{(t+2)} z^\cA N_\cA{}^\cB \del_\cB +\frac{2}{(t+2)} z^\cA N_\cA{}^\cB \del_\cB (z^\cC\del_\cC)
-\frac{K\cdot L}{(t+1)(t+2)}\left(z^\cA \del_\cA (z^\cB \del_\cB)-z^\cA \del_\cA\right)\ .
\label{5.7}
\ee
We can get rid of the $N$ terms if we restrict the differential operators to act on (germs of) functions which are homogeneous of degree $w$, \ie $z^\cA\del_\cA f= wf$. If we choose $w=-\frac{t}{2}$ we find
\be
\cD_K\cD_L= \cD_M+\half\cD_{[K,L]}-\frac{t}{4(t+1)} K\cdot L\ .
\la{5.8}
\ee
This agrees with \cite{Eastwood:2007} for the case where the algebra is chosen to be $\gs\gl(n|0)=\gs\gl(n)$, $n\neq 2$ and where $K\cdot L=\frac{n}{2} <K,L>$, where $<K,L>$ denotes the Killing form.

In the supersymmetric case there are clearly special cases where the above formul{\ae}   can have singularities that indicate indecomposable representations.  In the first instance, the traceless tensor $M$ is not well-defined if $(t+2)=0$, \ie $N=6$ in the $D=4$ superconformal algebra. There is also a problem if $(t+1)=0$ ($N=5$). This corresponds to the fact that the double-delta term in a second rank tensor is traceless by itself in this case. Finally, we note that the scalar term disappears if $t=0$, \ie $N=4$. This can be interpreted as a manifestation of the fact that the Killing form is itself zero in this case. Moreover, one can check explicitly that $K\cdot L$ given in \eq{4.7} also vanishes for $N=4$.

\section{Algebras and ideals}

In \cite{Eastwood:2002su} it was shown that, for the Laplacian in Euclidean space, the symmetries determined by conformal Killing tensors form an associative algebra given by the tensor algebra $T(\gg)$ of the conformal algebra $\gg$, which is $\gs\go(1,D+1)$ in this case, modulo its Joseph ideal which is generated by the relation
\be
K\otimes L=K\circledcirc L+ \half [K,L] - \l <K,L>,
\la{7.1}
\ee
where the first term is the Cartan product and the angle-brackets denote the Killing form. The symmetry involving a leading term with $n$ derivatives is determined the CKT given by the $n$-fold Cartan product of $\gg$. The constant $\l$ is unique (for $(D+2)>4$).

The notion of the Joseph ideal can be discussed quite generally for complex simple Lie algebras $\gg$, and in each case there is an ideal determined by a relation of the form \eq{7.1}. Moreover uniqueness of the constant $\l$ can be established by means of special tensors for each of the algebras $\gs\gl(n),\,n>2$, $\gs\go(n),\, n>4$ and $\gs\gp(n),\,n>1$ \cite{Eastwood:2007}. This method is very straightforward and we briefly review it here. 

The idea is that one constructs a tensor $S\in \gg\otimes\gg\otimes\gg$, starting from an arbitrary element $T\in\gg$. For each of the three series we can represent $T$ as a tensor on $V\otimes V$ ($V\otimes V^*$  for $\gs\gl(n)$), where $V$  carries the fundamental representation of $\gg$, so that $S$ has three pairs of indices. We then require $S$ to be antisymmetric under the interchange of the first two pairs, \ie $S\in\wedge^2\gg\otimes \gg$, to have no component in the Cartan product $\gg\circledcirc\gg$  of the second and third pairs of indices, and to have a non-zero projection onto the identity on the last two pairs. One can then reduce $S$ modulo the ideal generated by the relation \eq{7.1} in two different ways, \ie on the first two pairs and the last two pairs, to obtain a condition on $\l$. This condition says that there is a unique value of $\l$ such that the quotient of $T(\gg)$ modulo the ideal is infinite-dimensional. For all other values $T$ (and hence $\gg$) is itself in the ideal. The simplest example is perhaps the symplectic one, $\gg=\gs\gp(n)$, which we briefly summarise.

$S$ is taken to be
\be
S^{ab,cd,ef}=\left(8(\o^{ae}\o^{bf}T^{cd}-\o^{ac}\o^{be}T^{df})\right)-\left((ab)\leftrightarrow (cd)\right)\ ,
\la{7.2}
\ee
where each pair of indices is understood to be symmetrised. The indices run from $1$ to $2n$, $T$ is symmetric, and $\o$, the symplectic two-form, is antisymmetric. $S$ is clearly antisymmetric under the interchange of $ab$ with $cd$, and it is not difficult to verify that 
$S^{ab,(cd,ef)}=0$. This means that the Cartan product part on the last two pairs of indices vanishes. Reducing $S$ modulo \eq{7.1} on the first four indices and separately on the last four indices yields
\be
S\simeq -4(n-1)(n+1) T\simeq -2(n-1)(n+1)T+32\l(n-1)(n+1)^2 T\ ,
\la{7.3}
\ee
so that we must take $\l=-\frac{1}{16(n+1)}$ in order to avoid $T$ being in the ideal.

There are similar constructions for the other two cases. 

The above discussion can be taken over to  the supersymmetric case. It was given for $\go\gs\gp$ and $\gs\gp\go$ in \cite{css} and for $\gs\gl$ in \cite{Barbier}.
It turns out that the special tensor argument gives the same answers provided that traces are replaced by supertraces, although there are some special values of the dimensions of the algebras for which the discussion does not hold. We shall now go through each case in turn.
\subsection*{$D=3$}

For $D=3$ the superconformal algebras are $\gs\gp\go(2|N)$. An $n$th rank SCKT is given by a totally symmetric rank $2n$ tensor and can be represented by a one-row Young tableau with $2n$ boxes. All of these representations are irreducible so that there are no indecomposability problems. We have shown above that the leading term of an $n$th rank symmetry is given by such a tensor. The Joseph ideal is determined by considering the composition of two first-order symmetries and the value of $\l$ is indeed that given in \cite{css}. So it seems that there are no problems in this case.

\subsection*{$D=4$}

The situation is quite different for $D=4$. The product of two differential operators on super-twistor space has the value of $\l$ one would expect by taking the purely even case and replacing the trace by the supertrace, and this is also what one finds from the earlier superspace discussions. However, the special tensor argument of \cite{Barbier} breaks down for $N=2,4,6$. The cases of most interest from a physical point of view are $N=1,2,3,4$, so that this result is rather unfortunate especially for analytic superspace. The reason this arises is that one finds factors of $(t-2)$ and $t$ in the reduction of $S$, where $t=4-N$, which vanish for $N=2,4$ and because the Cartan product is problematic for $N=6$ since the second-rank symmetric, traceless representation is indecomposable in this case. For the special tensors, if one ignores the zeroes, which cancel out, one does get the expected values for $\l$ which coincide with those we have calculated in analytic superspace. In fact, one could take the view that the special tensor formul{\ae}  are valid for arbitrary values of $t$, so that one can continue into the complex $t$-plane, carry out the calculation there and then return to integral values of $t$ at the end, as in dimensional regularisation.  Further more, for all $N<4$ one does not encounter indecomposability problems for higher-order symmetries. On the other hand, for increasing $N>4$, there are always indecomposability problems. We shall not discuss these further here on the grounds that they are not expected to be of physical interest.

It therefore seems reasonable to propose that the algebra of symmetries of the super-Laplacian operators are given by $T(\gg)$ modulo the Joseph ideal for $N=1,2,3$. $N=4$ is a special case because the simple superconformal algebra in this case is projective, \ie $\gp\gs\gl(4|4)$.  Note that, in the super-twistor discussion, it was assumed that $K_{\cA}{}^{\cB}$ was traceless, and as the function germs that we used had weight zero, the discussion given there was indeed about $\gp\gs\gl(4|4)$. On the other other hand, an $N=4$ SCKV on analytic (or Minkowski) superspace contains an additional component corresponding to the abelian algebra $\gu(1)_Y$, so that one actually has a representation of $\gp\gg\gl(4|4)$, and this is also true for the symmetry operator $\cD_K$. For higher-rank SCKTs, however, the solutions to the differential equations defining them in analytic (or Minkowski) superspaces are irreducible tensors which correspond to representations of $\gp\gs\gl(4|4)$.  There are therefore no indecomposability problems for higher-order symmetries. We are therefore led to the conclusion that the algebra of symmetries for $N=4$ is that of the tensor algebra of $\gp\gs\gl(4|4)$ modulo its Joseph ideal together with an additional element corresponding to $\gu(1)_Y$.

\subsection*{$D=6$}

For $D=6$ the superconformal algebras are $\go\gs\gp(8|N)$. The SCKTs in this case are given, as in the bosonic case, by two-row Young tableaux which are in addition totally traceless. There is a problem defining the Cartan product for $N=3$ because the Cartan product of two elements of the algebra is obstructed by an indecomposability issue. There are also indecomposability problems for higher-rank SCKTs for $N>3$. However, the main cases of physical interest are $N=1,2$. In these two cases the values of $\l$ we have calculated are in agreement with those given in \cite{css}, and absence of any indecomposability problems would seem to indicate that we should be able to apply the arguments of \cite{Eastwood:2002su} to show that the algebra of symmetries is indeed given by the quotient of the tensor algebra of $\gg$ by the Joseph ideal.

\section{Concluding remarks}

In this paper we have seen that super-Laplacian operators can be defined in analytic superspaces for $D=3,4,6$, and we have shown that rank-$n$ generalised symmetries of super-Laplacians are determined by rank-$n$ SCKTs. In all cases this result holds even if the SCKT in question corresponds to  a reducible but indecomposable representation of the relevant superconformal algebra. Generically, it is not completely clear what the right definition of the algebra of symmetries is, precisely because of the role played by these reducible but indecomposable representations, but the situation is clearer for the cases of most physical interest, \ie $N\leq 2, D=6; N\leq 4, D=4; N\leq 8, D=3$. In these cases we have argued that the notions of Cartan product and  Joseph ideal are well-defined and therefore that the symmetry algebras can be characterised as quotients of the appropriate tensor algebras modulo their Joseph ideals. Indeed, for $D=3$ it may even be the case that one can make such a construction for arbitrary $N$.
\section*{Acknowledgements}
UL gratefully acknowledges the hospitality of the theory group at Imperial College, London,  partial financial support by the Swedish Research Council through VR grant 621-2013- 4245, as well as by the EPSRC programme grant EP/K034456/1. 
\section*{Appendix:\ signs}
Let $V$ be a super vector space with dual $V^*$ and bases $e^A$ and $e_A$ respectively, where $A=(a,\a)$ is super-index with $(a,\a)$ being respectively even and odd. We shall choose the summation convention running from NW to SE, and suppose that all tensor components have the Grassmann parity indicated by their indices, although one could easily allow for intrinsic Grassmann parities as well. We set
\be
V\ni X=X^A e_A \ ; \qquad V^*\ni \o=e^A\o_A\ .
\ee
In order to avoid signs for higher-rank tensors, the indices should be summed in pairs starting from the inside and working to the outside. For example, a tensor $T$ of type (2,2) \ie an element of  $V^*\otimes V^*\otimes V\otimes V$ should be written
\be
T=e^B\otimes e^A T_{AB}{}^{CD} e_D\otimes e_C\ .
\ee
The convention we use is that all of our equations are tensorial so that we do not need to write out the signs explicitly. A simple example is the super-trace of a (1,1) tensor $T_A{}^B$. We write this as $T_A{}^A$ with the understanding that the indices being summed in the ``wrong'' order means that a factor $(-1)^A$ is understood, where this factor is $+1$ for $A=a$ and $-1$ for $A=\a$. In particular taking the trace over the unit tensor $\d_A{}^B$ will give $\d_A{}^A=t$, where $t$ is the difference between the dimension of the even and odd parts of the space in question. Summations of this type occur in the definition of SCKTs, for example. For graded (anti)-symmetrisations the correct signs are understood. Thus 
\be
(AB)=\half (AB+(-1)^{AB} BA) \ :\qquad [AB]=\half (AB-(-1)^{AB} BA) \ .
\ee
If $T_{AB,C}$ is graded antisymmetric on its first two indices then 
\be
T_{[AB,C]}=\frac{1}{3}( T_{AB,C} + (-1)^{C(A+B)}T_{CA,B} + (-1)^{A(B+C)}T_{BC,A})\ .
\ee
In other words one (anti-)symmetrises in the usual way but then adds signs so that the indices that are in the ``wrong'' order are corrected. But we do not need to write these signs out explicitly. In short, for all the super manipulations that we carry out in this paper we can ignore the Grassmann signs because all expressions are tensorial; the only change from the purely even case is that traces are replaced by 
supertraces. This is the case for the calculations carried out in section 3, for example.

On the other  hand, when one wishes to look at the different components of super equations it is necessary to get the signs right, for example in the discussion of the minimal multiplets in section 2.3. Consider, for example, the super-Laplacian in $D3$ analytic superspace \eq{2.16}. We wrote it as
\be
\D_{AB,CD}=\del_{[A[C}\del_{B]D]}\ .
\ee
 This is a tensor operator in $(V^*)^{\otimes 4}$. Since it is a product of two derivatives only one pair of anti-symmetrisation brackets is required. The correct formula is then
 \be
 2\D_{AB,CD}=(-1)^{BC}\del_{AC}\del_{BD}-(-1)^{AB+AC} \del_{BC}\del_{AD}\ .
 \ee
 The Grassmann factor in the first term on the right is because $B$ and $C$ are in the wrong order for the tensor structure on the left. The factor in the second term is the graded anti-symmetrisation factor together with another factor because $A$ is the ``wrong'' side of $C$.

Finally, in $D=4$ analytic superspace there are two types of index, primed and unprimed, so to avoid sign factors one should choose an ordering of the primed and unprimed indices, \eg unprimed to the left for covariant indices and to the right for contravariant ones, such as $T_{AA'}{}^{B'B}$. However, we do not need to do this explicitly for the reasons stated above.

\end{document}